\documentclass[aps,twocolumn,groupedaddress,floatfix,superscriptaddress]{revtex4}
\usepackage{graphicx}
\usepackage{amsmath}
\usepackage{amstext}
\usepackage{amssymb}
\usepackage{amsfonts}
\usepackage{amsbsy} 

\renewcommand{\v}[1]{\boldsymbol{#1}} 

\newcommand{\Ref}[1]{Ref.~\cite{#1}}
 
\newcommand{\<}{\langle} 
\renewcommand{\>}{\rangle}

\newcommand{\si}{\sigma}

\newcommand{\bpm}{\begin{pmatrix}}
\newcommand{\epm}{\end{pmatrix}}
\newcommand{\bmm}{\begin{matrix}}
\newcommand{\emm}{\end{matrix}}

\begin{document}
\bibliographystyle{apsrev}

\title{Collective states of non-abelian quasiparticles in a magnetic field}

\author{Michael Levin}
\affiliation{Department of Physics, University of California, Santa Barbara, California 93106}
\affiliation{Department of Physics, Harvard University, Cambridge, Massachusetts 02138}
\author{Bertrand I. Halperin}

\affiliation{Department of Physics, Harvard University, Cambridge, 
Massachusetts 02138}

\date{\today}

\begin{abstract}
Motivated by the physics of the Moore-Read $\nu = 1/2$ state
away from half-filling, we investigate collective states of non-abelian 
$e/4$ quasiparticles in a magnetic field. We consider two types of 
collective states: incompressible liquids and Wigner crystals. In the 
incompressible liquid case, we construct a natural series of states
which can be thought of as a non-abelian generalization of the Laughlin 
states. These states are associated with a series of hierarchical 
states derived from the Moore-Read state - the simplest of which occur
at filling fraction $8/17$ and $7/13$. Interestingly, we find that the
hierarchical states are abelian even though their parent state is 
non-abelian. In the Wigner crystal case, we construct two candidate
states. We find that they, too, are abelian - in agreement with previous 
analysis.
\end{abstract}
\pacs{}

\maketitle

\section{Introduction}
The Moore-Read Pfaffian state \cite{MR9162} (or its particle-hole conjugate,
the anti-Pfaffian  \cite{LHR0706, LRNF0707})) is believed to be a good 
candidate for the observed $\nu = 5/2$ quantum Hall plateau. \cite{XPV0409} 
This possibility is particularly exciting since the quasiparticle 
excitations in this state carry non-abelian statistics. Much work has been 
devoted to understanding the  basic physical properties of the Pfaffian state. 
However, one aspect of the Pfaffian state has received less 
attention - the physics of this state away from half filling.

To understand the basic issue, consider an idealized, perfectly
clean FQH system whose ground state at half filling is the Pfaffian state. 
Suppose that the magnetic field is adjusted so that the filling fraction is 
slightly less than $1/2$. In this case, we expect that the ground state 
will be given by the Pfaffian together with a small but finite density of $e/4$ 
quasiholes. 

This unusual physical system motivates a general question: what kinds of 
collective states can arise from a finite density of non-abelian $e/4$ 
quasiholes in a magnetic field? This is the main subject of this paper. We 
consider two basic kinds of collective states. The first kind of state is a 
Wigner crystal of quasiholes. Such a state is natural if the 
dominant interaction between the quasiholes is the repulsive Coulomb force, and
the quasiholes are sufficiently dilute. Thus it is a good candidate for
a clean FQH system at filling $\nu$ very close to $5/2$.
The second possibility we consider is that the quasiholes form
an incompressible liquid. This is perhaps less relevant 
to a physically realistic $\nu = 5/2$ system, but it is interesting 
conceptually. In the context of the Pfaffian state, such an
incompressible liquid would correspond to a hierarchical quantum Hall
state - but unlike the usual FQH hierarchy construction, \cite{H8305,H8483} 
it would be built out of \emph{non-abelian} anyons.

One way that quasiholes can form an incompressible liquid is if they pair
to form charge $e/2$ abelian anyon boundstates, as has been proposed by
Bonderson and Slingerland. \cite{BS0823} These abelian anyons can 
then form Laughlin-like incompressible states. The result is a series of 
hierarchical states which carry non-abelian statistics similar to the 
Pfaffian state. 

In this paper, we construct a different kind of condensate - one made
up of individual non-abelian quasiholes, rather then abelian anyon molecules. 
The motivation for this is conceptual: it is not obvious what kinds of 
incompressible liquids can form from non-abelian anyons. Currently, we do not have good
reason to decide whether this type of condensate, the paired condensate
in \cite{BS0823}, or something entirely different is most likely in a
realistic $\nu = 5/2$ system. Our philosophy is rather that each of these
possibilities can occur in principle (depending on the details of the electron
interaction) and therefore merits consideration.

Our starting point is a model where the quasihole interaction is short range, two-body
and repulsive, and the quasiholes are in their lowest effective Landau level.
Our main result is that we construct a series of incompressible liquids that are
good candidate ground states for such a model. These states can
be regarded as a non-abelian generalization of the Laughlin states. \cite{L8395}
In the context of the Pfaffian state near half-filling, these states are associated
with a series of hierarchical FQH states at filling fractions $8m/(16m+1) = 8/17,16/33,...$ 
Interestingly, we find that these states are actually abelian quantum Hall 
states - even though they are derived from the non-abelian Pfaffian state.
We find that these states are distinct from the usual Jain states \cite{J8999} 
(or equivalently hierarchical states \cite{BW9045}) at filling fractions 
$8m/(16m+1)$ but are equivalent to hierarchical states derived from
the strong pairing $\nu = 1/2$ state. \cite{H8375} We also consider incompressible
liquids of $e/4$ quasiparticles. We again find a series of states - in
this case at filling fractions $(8m-1)/(16m-3) = 7/13, 15/29,...$. These states 
are also abelian, and distinct from the Jain states. They are equivalent to 
hierarchical states derived from the $331$ $\nu = 1/2$ state. \cite{H8375,HR8886}

In addition, we consider the problem of a Wigner crystal of quasiholes. The 
physics of this system is nontrivial due to the non-abelian statistics of the 
quasiholes. Even after the quasiholes localize at certain positions in space, 
there are still $2^{N_{\text{qh}}/2-1}$ nearly degenerate quasihole states 
coming from their non-abelian statistics. To fully specify the ground 
state, one needs to specify a state in this $2^{N_{\text{qh}}/2-1}$ 
dimensional Hilbert space. Previous analysis of an exactly soluble (but 
simplified) model yielded a particular ground state which, in the context
of the Pfaffian state, turned out to be an abelian state \cite{AK2006} 
(see also \cite{GS0603} and in the disordered case, \cite{RL0104}). 
More specifically, this state turned out to be in 
the same universality class as the strong pairing state. 
In this paper, we simply look for natural candidate states in the 
$2^{N_{\text{qh}}/2-1}$ dimensional Hilbert space. We find two of them, 
both abelian. One is in the strong pairing universality class, and one is 
in the $331$ universality class. While our approach is not well controlled 
like the analysis in \Ref{AK2006}, it has the advantage that it
can be generalized to a Wigner crystal of Read-Rezayi quasiholes \cite{RR9984}- 
a system which appears less amenable to the exact analysis of \Ref{AK2006}.

Currently, there is no experimental evidence for the above series of
hierarchical states. No plateaus have been found at any of the
filling fractions $8/17$, $7/13$, etc. Therefore, we feel the
main contribution of this paper is conceptual. We hope that our discussion
reveals some of the basic phenomena in the many-body physics 
of non-abelian anyons. 
 
The paper is organized as follows. In sections \ref{qmech}-\ref{incomp} 
we consider the idealized problem of a gas of point like non-abelian 
anyons in a magnetic field. We give a quantum mechanical description of 
this system in section \ref{qmech}, and we propose a series of
candidate incompressible states in section \ref{incomp}. In sections 
\ref{prophi}-\ref{jain} we apply these results to the Pfaffian state. We 
construct a series of hierarchical quantum Hall states built out of the 
non-abelian quasiholes and examine their properties. In section \ref{qpqh} 
we repeat the analysis for the non-abelian quasiparticle, and construct 
another series of hierarchical states. In sections \ref{hihi}-\ref{antipf} 
we consider higher level hierarchical states and states derived from
the anti-Pfaffian state. Finally, in section \ref{locqh} we apply our approach
to the problem of a quasihole Wigner crystal.

\section{Quantum mechanics of non-abelian anyons}
\label{qmech}
In this section, we address a preliminary question - how to formulate 
quantum mechanics for point-like non-abelian anyons. We focus on the case 
of non-abelian anyons with the same statistics as the quasihole in the 
Pfaffian state, as that is the case that is relevant to our later analysis. 
We also assume the particles are hardcore so as to avoid singularities 
when particles occupy the same point in space. 

To begin, consider a system of four such non-abelian anyons and 
assume that the anyons are at fixed positions $r_1, r_2, r_3, r_4$. 
Because of the non-abelian statistics of the anyons, the four anyon system 
can be in two states. Thus, the Hilbert space for this system is two 
dimensional, and the wave functions that describe the states are two 
component row vectors $(\Psi_1 ,\Psi_2)$. The Hamiltonian for such a 
system can be any $2 \times 2$ Hermitian matrix.

Now consider the case where the anyons are free to move. A natural guess 
is that the appropriate wave functions for this system are two component 
row vectors of \emph{functions}, $(\Psi_1(r_1,r_2,r_3,r_4) , 
\Psi_2(r_1,r_2,r_3,r_4))$ while the Hamiltonian is the usual kind of
differential operator. This is essentially correct, but we need to 
take account of the Berry phases associated with the exchange of the
anyons in some way. There are two ways to do this. One can either add a 
non-local Chern-Simons interaction to the Hamiltonian, or one can modify 
the Hilbert space so that the allowed wave functions are multi-valued. 

In this paper, we 
will use the second approach. We take our wave functions $(\Psi_1 , 
\Psi_2)$ to be multi-valued on the four particle configuration space
$\{\{r_1,r_2,r_3,r_4\}: r_i \neq r_j\}$ - or more 
accurately, single valued on a Riemann sheet covering this space. 
We impose a constraint on the wave functions analogous to the 
anti-symmetry constraint on fermionic wave functions. The constraint is 
defined as follows. Recall that for every possible braid or exchange of 
the four anyons, there is a corresponding $2 \times 2$ unitary matrix 
$U_{\alpha \beta}$ which describes the effect of this exchange on the two 
degenerate states: $|\alpha,ex\> = \sum_\beta U_{\alpha \beta} |\beta\>$. 
We constrain our wave functions to have the 
property that under such an exchange, the $(\Psi_1 , \Psi_2)$ transform 
according to this unitary matrix: $\Psi_{\alpha,ex} = 
\sum_\beta \Psi_\beta U_{\beta \alpha}^*$. Also, since the anyons are 
hardcore particles, we impose the boundary condition that $\Psi_{1,2} 
\rightarrow 0$ as $r_i \rightarrow r_j$.

In the multi-valued wave function approach, the Berry phases are taken 
care of by the choice of Hilbert space. The Hamiltonian for a $4$ anyon 
system is therefore no different from a $4$ boson or $4$ fermion system 
(at the level of a differential operator). For example, the appropriate 
Hamiltonian for describing $4$ anyons with quadratic dispersion, charge $q$, 
mass $m$, and subject to a uniform magnetic field $B$ in the $\hat{z}$ 
direction is simply
\begin{equation}
H_0 = \sum_{i=1}^4 \frac{1}{2m} \left(\frac{1}{i} \vec{\nabla}_{r_i} - q \vec{A}(r_i) \right)^2
\label{anyonH}
\end{equation}
where $\vec{\nabla} \times \vec{A} = B$. Interactions (beyond the hardcore constraint) 
can be introduced by including a potential energy term: $H = H_0 + 
V(r_1,r_2,r_3,r_4)$. 

It is worth mentioning that in general non-abelian systems, the potential 
energy term can be more complicated. Indeed, the most general potential 
energy term for a $4$ anyon system is described not by a scalar, but 
rather by a $2 \times 2$ matrix $V_{\alpha \beta}(r_1,r_2,r_3,r_4)$, 
$\alpha, \beta = 1,2$. This $2 \times 2$ matrix is multi-valued, and 
transforms like $V_{ex} = U^{-1} V U$ when two particles are exchanged. 
Physically, such a term describes an energy splitting between the two 
possible fusion outcomes of the four particle system. In generic 
non-abelian anyon systems, we expect such a splitting to be 
present, though it will be exponentially small in the particle separation 
divided by a microscopic length scale (which is on the order of
the magnetic length in the 
case of quantum Hall systems). Because of this exponential suppression, 
and because of the complexity of including such terms, in this paper we 
will restrict our attention to scalar potential energy interactions.

We can now easily generalize these results to a system of $2N$ anyons.
In the general case, the Hilbert space is described by row vectors 
$(\Psi_1, ... , \Psi_{2^{N-1}})$ of $2^{N-1}$ functions of $2N$ variables, which 
transform under appropriate unitary matrices when two particles are 
exchanged. We again impose the boundary condition that $\Psi$ vanishes as 
$r_i \rightarrow r_j$. As for the Hamiltonian, this is completely analogous 
to the $4$ anyon case.


\section{Incompressible states of non-abelian anyons in a magnetic field}
\label{incomp}
In this section, we construct a series of incompressible states of non-abelian
anyons in a magnetic field. These states apply to a particular kind 
of anyon - anyons with the braiding statistics of the quasiholes in 
the Pfaffian state. (We will also take the anyons to be hardcore to avoid 
technical complications). We argue that these states are good candidate 
ground states in the case where the anyon interactions are 
short-range, two-body, and repulsive and the anyons are in their lowest effective 
Landau level. 

We would like to mention that our model of short-range, two-body, repulsive 
interactions is motivated more by its simplicity than its realism. Indeed, 
in the physical case of the Pfaffian state at filling $\nu = 5/2 - 
\epsilon$, we expect that the quasihole interaction to be somewhat 
different. One can distinguish three parts to the interaction in this case: 
(1) There may be a long range Coulomb interaction, falling off as $1/r$, 
whose strength is determined by the quasiparticle charge $e^*$ and the 
dielectric constant of the material. In principle this piece could be 
absent, for example, if there is a nearby metallic gate which cuts off the 
Coulomb interaction at long distances. (2) At short distances there
will be a portion of the interaction that we may describe by an effective 
potential, whose details are not known to us. (3) There will be interaction
terms that depend on the fusion channel and cannot be represented by a 
simple position dependent potential. As we discussed in the previous 
section, these interactions have an energy scale $E \sim 
E_0 \exp(-r/\xi)$ for some parameters $E_0, \xi$. The precise values of $E_0, \xi$ are
unknown but $\xi$ is typically of the same 
order as the magnetic length $l^*$ for quasiparticles of charge $e^* = e/4$, while
$E_0$ is of the same order as the energy gap $\Delta$, at least in some cases. In this paper, we ignore 
interactions of type (3), effectively considering the limit $\xi \ll l^*$ or $E_0 \ll \Delta$. We also treat 
the rest of the interaction as a short range repulsion. While neither of these 
assumptions are realistic, this model is at least a simple starting point.

We begin by analyzing the Landau level physics of our non-abelian anyons. 
Consider the Hamiltonian for $2N$ non-interacting anyons with charge $q$ in 
a magnetic field $B$. In the general case where the anyons do not 
necessarily have quadratic dispersion, the appropriate Hamiltonian is of the form
\begin{equation} 
H_0 = \sum_i F \left((\frac{1}{i} \vec{\nabla}_{r_i} - q \vec{A}(r_i))^2\right)
\end{equation}
where $F(x)$ is some increasing function for $x > 0$, not necessarily linear, and 
$\vec{\nabla} \times \vec{A} = B$. Independent of the details of $F$, the lowest energy 
states of this Hamiltonian are given by solutions 
$(\Psi_1, ..., \Psi_{2^{N-1}})$ of
\begin{equation}
\left(\frac{1}{i} \vec{\nabla}_{r_i} - q \vec{A}(r_i)\right)^2 \Psi_j = |q| B \Psi_j
\label{lll}
\end{equation}
(This follows from the fact that the differential operator $(\frac{1}{i} 
\vec{\nabla}- q \vec{A})^2$ has lowest eigenvalue $|q|B$).
These low energy states are highly degenerate, and all have energy $2N 
F(|q|B)$. Physically, these states correspond to the case where all $2N$ 
anyons are in their lowest effective Landau level. If the gap separating 
these lowest Landau level states from higher energy states is larger then the 
interaction energy scale, then we can restrict our Hilbert space to these states,
replacing the anyon interaction with an appropriate lowest Landau level
interaction. Our problem then reduces to understanding 
which lowest Landau level states are favored by the interactions.
This is the limit we will consider here.

It is convenient to write down the lowest Landau level states more explicitly. 
Let us focus on the case $qB < 0$, and choose the circular gauge, $\vec{A} = B 
(x \hat{y} - y \hat{x})/2$. Using complex coordinates $w = x + iy$, the solutions
to (\ref{lll}) can be written as
\begin{equation}
(\tilde{\Psi}_1(\{\bar{w}_i\}), ..., \tilde{\Psi}_{2^{N-1}}(\{\bar{w}_i\} 
) \cdot e^{-\frac{1}{4 l^2} \sum_i |w_i|^2}
\end{equation}
where $(\tilde{\Psi}_1, ...,\tilde{\Psi}_{2^{N-1}})$ are anti-analytic in 
$w_1,...,w_{2N}$, and $l^2 = 1/|q|B$. Since the particles in question are 
non-abelian anyons, these functions need to be defined on an appropriate 
Riemann sheet and satisfy the transformation law described above 
under particle exchange. 


We can now proceed to the problem of constructing candidate incompressible 
states. Following the logic of the Laughlin construction, we look for 
states which are (a) in the lowest Landau level and (b) have high order 
zeros when two particles approach each other. Here, the second condition 
comes from our assumption that the anyons have short-range, two-body, 
repulsive interactions. Given the above parametrization of the lowest Landau 
level, this reduces to the problem of finding
a row vector $(\tilde{\Psi}_1, ... , \tilde{\Psi}_{2^{N-1}})$ of 
\emph{anti-analytic} functions satisfying the appropriate transformation 
law under particle exchange, and having high order zeros when 
$w_i \rightarrow w_j$.

It is not obvious how to write down a collection of anti-analytic 
functions with the appropriate transformation properties, much
less one with high order zeros when two coordinates coincide. 
However, the conformal field theory ansatz for 
constructing trial fractional quantum Hall states provides a natural 
solution to both of these problems. The idea is to define a trial anyon 
wave function using a correlator from conformal field theory. \cite{MR9162} In 
particular, consider the correlator
\begin{displaymath} 
\<\bar{\si}(\bar{w}_1)e^{i\sqrt{2m+\frac{1}{8}}\bar{\phi}(\bar{w}_1)} \cdots 
\bar{\si}(\bar{w}_{2N})e^{i\sqrt{2m+\frac{1}{8}}\bar{\phi}(\bar{w}_{2N})}\>
\end{displaymath}
defined in 
a CFT which is a product of a chiral Ising theory \cite{FMS97, MR9162} and a chiral
boson theory. Here,
$m$ is a positive integer, $\bar{\si}$ is the spin operator in the chiral Ising 
theory and $\bar{\phi}$ is a chiral boson operator in the chiral 
boson theory with normalization convention 
\begin{equation}
\< e^{i\sqrt{2m+\frac{1}{8}}\bar{\phi}(\bar{w}_{1})} e^{-i\sqrt{2m+\frac{1}{8}}\bar{\phi}(\bar{w}_{2})}\>
= (\bar{w}_1 - \bar{w}_2)^{-2m-\frac{1}{8}}
\end{equation}
This correlator has $2^{N -1}$ different conformal 
blocks - which we can label by $\alpha = 1,...,2^{N-1}$. Thus, we can 
define $2^{N-1}$ anti-analytic functions,
\begin{equation}
\tilde{\Psi}_{\alpha}^m(\bar{w}_1,...,\bar{w}_{2N}) =
\<\bar{\si}(\bar{w}_1)e^{i\sqrt{2m+1/8}\bar{\phi}(\bar{w}_1)} \cdots\>_{\alpha}
\end{equation}

It is not hard to see that these anti-analytic functions satisfy the 
appropriate transformation law under particle exchange. Indeed, this follows 
from the monodromy properties of the above correlation 
function (e.g. its transformation law under analytic continuation). 
\cite{FMS97, MR9162} In addition, the correlation functions naturally satisfy the 
hardcore boundary condition that $\tilde{\Psi} \rightarrow 0$ when 
$w_i \rightarrow w_j$. Thus, these states - which we will denote by 
$\Psi^m$ - are legitimate many-body anyon states in the lowest Landau level.

For example, evaluating the correlator in the case $N = 2$, one finds
\begin{equation}
\tilde{\Psi}_{1,2}^m(\bar{w}_1,\bar{w}_2,\bar{w}_3,\bar{w}_4) = 
(\sqrt{\bar{w}_{13}\bar{w}_{24}} \pm
\sqrt{\bar{w}_{14}\bar{w}_{23}})^{\frac{1}{2}}\prod_{i <
j} \bar{w}_{ij}^{2m}
\end{equation}
where $w_{ij} = w_i - w_j$. One can see that when particles 
$1$ and $2$ are exchanged in the counterclockwise direction, 
$(\tilde{\Psi}_1 , \tilde{\Psi}_2) \rightarrow 
(\tilde{\Psi}_1 , -i \tilde{\Psi}_2)$. This is exactly the unitary 
transformation we expect for a particle exchange. Moreover, one can see 
that $\tilde{\Psi}_1, \tilde{\Psi}_2$ vanish as $w_i \rightarrow w_j$.

In addition to being in the lowest Landau level, the $\Psi^m$ also have
the desired high order zeros when particles approach each other. Indeed, 
one can see that $\Psi^m$ has a $2m^{th}$ order zero when $w_i \rightarrow 
w_j$ (More specifically, $\Psi^m$ vanishes like $(\bar{w}_i - 
\bar{w}_j)^{2m}$ when $i, j$ are in the identity fusion channel, and 
$(\bar{w}_i - \bar{w}_j)^{2m+1/2}$ when $i, j$ are in the $\psi$ fusion 
channel). 

Since the $\Psi^m$ have the desired properties (a), (b), they appear to be
good candidate ground states for short range, two-body, repulsive 
interactions. However, to complete the story, it would be good to find a 
specific interaction for which they are the exact ground states. Such a model
interaction is easy to construct. Indeed, let $V = 
V(r/a)$ be a short-range repulsive two-body interaction with range $a$. 
The model interaction we consider is simply $V$ in the limit of small $a$ 
(e.g. $a \ll l$). (Note that this is the same as the model interaction 
which stabilizes the Laughlin states).

To see why $\Psi^m$ is the ground state for this interaction, note that the 
interaction energy of $\Psi^m$ is very small for small $a$, due to the 
presence of the $2m^{th}$ order zeros:
\begin{equation}
\<\Psi^m| \sum_{i < j} V(\v{r}_i-\v{r}_j) | \Psi^m\> \propto \left( \frac{a}{l} \right)^{4m+2}
\end{equation}
Also, it is not hard to see that $\Psi^m$ scales like $w_i^{2m(2N-1) + 
(N-1)/4}$ as $w_i \rightarrow \infty$. This means that the highest single 
particle angular momentum in $\Psi^m$ is $L_{max} = 2m(2N-1) + (N-1)/4$. We 
believe that any other state with maximum angular momentum
$L_{max}$ has at least one lower order zero. We have
not proven this statement, but it appears to hold for small systems.
For example, in the $N = 2$ case, one
can prove that $\Psi^m$ is the unique state with maximum 
angular momentum $L_{max} = 6m + 1/4$ and $2m^{th}$ order zeros. Assuming
this is true is general, it follows that all other states on an
appropriately sized disk or sphere have an interaction energy which goes
to zero slower than $(a/l)^{4m+2}$. Thus, in the limit $a
\rightarrow 0$, $\Psi^m$ will be the unique ground state.

We cannot make any analytic arguments for the existence of a finite gap. 
This must be checked numerically. However, given the analogy to the Laughlin 
states, a gap seems likely for small $m$. On the other hand, when $m$ is large,
the system may be gapless since in that case $\Psi^m$ may describe a Wigner
crystal rather than a uniform density liquid.

Assuming that $\Psi^m$ is gapped for small $m$, it is necessarily 
translationally invariant, and thus an incompressible liquid. 
We have therefore accomplished our goal of constructing
incompressible liquid states of non-abelian anyons in a magnetic
field. The filling fractions of these states are given by
\begin{equation}
\nu_{anyon} = \lim_{N \rightarrow \infty} \frac{2N}{L_{max}} = \frac{1}{2m+1/8}
\label{anyfill}
\end{equation}

\section{Properties of the hierarchical states}
\label{prophi}
We now return to the physics of the Pfaffian state away from half filling.
We investigate the hierarchical quantum Hall states $\Psi^m_{hi}$ 
that would result if the quasiholes formed one of the $\Psi^m$ states.

To begin, assume we have a FQH system whose ground state at half filling is 
the Pfaffian state, and suppose we increase the magnetic field,
so that the system nucleates a finite density of quasiholes. Since these
quasiholes feel a magnetic field with $e^* B < 0$, and carry the 
appropriate non-abelian statistics, they could in principle form one of 
the incompressible liquids described by $\Psi^m$. 

In practice, these states may not be realized. As we mentioned earlier, 
our model of point-like non-abelian anyons with short-range, two-body, 
repulsive interactions is not a good description of a realistic FQH system 
- especially when the quasihole density is high. On the other hand, when 
the quasihole density is very low, it is likely that the quasiholes will 
form a Wigner crystal. Nevertheless, it is still interesting conceptually
to understand the properties of these hypothetical hierarchical states $\Psi^m_{hi}$.

First, let us compute the filling fraction for these states. Recall 
that the anyon filling fraction in $\Psi^m$ is $\nu_{anyon} = 1/(2m+1/8)$. 
Since the anyons in this case are quasiholes with charge $e/4$, the 
corresponding electron filling fraction in $\Psi^m_{hi}$ is 
\begin{equation}
\nu = \frac{1}{2} - \frac{1}{16} \cdot \nu_{anyon} = \frac{8m}{16m+1}
\end{equation}
In particular, the simplest hierarchical state (with $m=1$) occurs at $\nu 
= 8/17$.

Next, we write down wave functions for these states in terms of the 
original electron coordinates $\{z\} \equiv \{z_1,...z_{2M}\}$. The logic is 
similar to the usual abelian hierarchical construction but it is worth 
repeating here, given the unfamiliar context. In the previous section, 
we thought of the wave function $\Psi^m_\alpha(\{\bar{w}\})$ as defining a 
state in a Hilbert space for point-like non-abelian anyons. This state was 
given by
\begin{equation}
|\Psi^m\> = \int dw \sum_\alpha 
\Psi^m_\alpha(\{\bar{w}\})\cdot |\{w\},\alpha\>
\label{hieq1}
\end{equation}
where $|\{w\},\alpha\>$ denotes the state in which the non-abelian anyons 
are at positions $\{w\}$ and in fusion state $\alpha$. Now we consider the
case where the non-abelian anyons are actually quasihole excitations in the 
Pfaffian state. In this context, the Hilbert space for non-abelian 
anyons is actually a \emph{subspace} of the full electron Hilbert 
space - the subspace spanned by $2N$ quasihole states. Each state 
$|\{w\},\alpha\>$ corresponds to some linear combination of electron basis 
states $|\{z\}\>$:
\begin{equation}
|\{w\},\alpha\> = \int dz \ \Phi_{\alpha,qh}(\{z\};\{w\}) \cdot |\{z\}\>
\label{hieq2}
\end{equation}
The coefficient in this expansion, $\Phi_{\alpha,qh}$, is the electron wave 
function for fixed quasihole positions $\{w\}$ and fusion state $\alpha$.

Combining (\ref{hieq1}) and (\ref{hieq2}), we see that $|\Psi^m\>$ corresponds
to the hierarchical state 
\begin{equation}
|\Psi^m_{hi}\> = \int dw dz \sum_\alpha \Psi^m_\alpha(\{\bar{w}\})  
\Phi_{\alpha,qh}(\{z\};\{w\}) \cdot |\{z\}\>
\end{equation}
Equivalently, in wave function notation, 
\begin{equation}
\Psi^m_{hi}(\{z\}) = \int dw \sum_\alpha \Psi^m_\alpha(\{\bar{w}\})
\Phi_{\alpha,qh}(\{z\};\{w\})
\label{wf}
\end{equation}

To proceed further, we use the explicit form for the quasihole wave functions 
$\Phi_{\alpha,qh}$. Like the $\Psi^m$, they can be written as
$\Phi_{\alpha,qh} = \tilde{\Phi}_{\alpha,qh} \cdot e^{-1/4 l^2 \sum_i |z_i|^2}$
where $\tilde{\Phi}_{\alpha,qh}$ is a CFT correlator. \cite{MR9162} 
Specifically,
\begin{align}
& \tilde{\Phi}_{\alpha,qh}(\{z\};\{w\}) =  \nonumber \\
&\<\si(w_1)e^{\frac{i}{\sqrt{8}}\phi(w_1)} \cdots
\psi(z_1)e^{i\sqrt{2}\phi(z_1)} \cdots\>_{\alpha}
\end{align}

Combining the two correlators into one, the sum in (\ref{wf}) can be 
simplified to
\begin{align}
&\sum_\alpha \Psi^m_\alpha(\{\bar{w}\}) \Phi_{\alpha,qh}(\{z\};\{w\}) = 
\nonumber \\ 
&\<\si(w_1,\bar{w}_1)e^{\frac{i}{\sqrt{8}}\phi(w_1) + 
i\sqrt{2m+\frac{1}{8}}\bar{\phi}(\bar{w}_1)}\cdots \psi(z_1)e^{i\sqrt{2}\phi(z_1)} 
\cdots\> \nonumber \\
&\cdot e^{-\frac{1}{4 l^2} \sum_i |z_i|^2 - \frac{1}{16 l^2} \sum_i |w_i|^2}
\label{sumid}
\end{align}
where $\si(w,\bar{w})$ is the spin field in the \emph{non-chiral} Ising model.
\cite{FMS97}

This simplification is useful because it expresses the wave function 
$\Psi^m_{hi}$ in terms of a single CFT correlator and thus allows us 
to use conformal field theory to analyze the state. \cite{MR9162} This 
conformal field theory approach is very powerful and allows us to quickly
extract the universal properties of the quasiparticles and edge 
excitations. However, it is worth mentioning that the CFT approach
is more a series of conjectures then a rigorous method. It is known to work
in some cases (such as the Laughlin state and Moore-Read state \cite{R0908}) 
and is believed to work in many others, but it has not been 
proven in generality. Therefore, the results below - like the existence 
of gap - need independent (numerical) verification.

We begin with an analysis of the quasiparticle spectrum. According to the 
conformal field theory approach, quasiparticle excitations can 
be constructed by inserting operators $\mathcal{O}(z_0)$ into the above
correlation function (\ref{sumid}). The operators $\mathcal{O}$ can be 
arbitrary except that they must be local with respect to the electron 
operator $\psi(z)e^{i\sqrt{2}\phi(z)}$ and the quasihole operator 
$\si(z,\bar{z})e^{i(\sqrt{1/8}\phi(z) + 
\sqrt{2m+1/8}\bar{\phi}(\bar{z}))}$ - that is, their correlation functions
with these operators must be single-valued. (This requirement comes 
from the fact that the excited state wave function must be single-valued 
in the electron and quasihole coordinates). The second 
part of the conformal field theory conjectures is that, in the 
thermodynamic limit, the Berry phase associated with exchanging two 
quasiparticle excitations with coordinates $z_0,z_1$ is exactly given by 
the monodromy of the associated correlation function 
$\<\mathcal{O}(z_0)\mathcal{O}(z_1)...\>$ under
such an exchange (e.g. the phase associated with analytically continuing
the correlation function along a path exchanging $z_0,z_1$). This is
known to be correct in the Laughlin case - where one can explicitly 
calculate the Berry phase using the plasma analogy - but it has not been 
established in more complicated cases such as the one in question.

Assuming that these conjectures can be applied in this case, we simply 
need to look for operators $\mathcal{O}$ which are local with respect to 
the electron and quasihole operator. We can then define an 
equivalence relation on these operators by setting $\mathcal{O}_1 \equiv 
\mathcal{O}_2$ if correlation functions involving $\mathcal{O}_1 
\mathcal{O}_2^{-1}$ and any other allowed operators are single-valued. 
The resulting equivalence classes should then be in one-to-one 
correspondence with topologically distinct quasiparticle 
excitations.

We begin with the set of operators of the form 
$\{\chi e^{i a \phi + i b \bar{\phi}}\}$ where $\chi = 1, \psi, \si, \mu$ and
$a,b$ are real numbers (in principle, one should consider even more general
operators with spatial derivatives, etc., but these
do not appear to give new equivalence classes). Imposing locality with respect to the electron
and quasihole operator, we find the following list of allowed operators,
or more accurately, equivalence classes of operators:
$\{\chi e^{i(n_1\phi/\sqrt{2}+(n_2+n_1/4)\bar{\phi}/\sqrt{2m+1/8})}\}$ 
where $n_1$ is an integer or half 
integer depending on whether $\chi = 1,\psi$ or $\chi= \si,\mu$ and $n_2$ 
is integer or half-integer depending on whether $\chi = 1,\si$ or $\chi = 
\psi,\mu$. One can check that all of these operators can be generated
from the operator $\psi e^{i\bar{\phi}/2\sqrt{2m+1/8}}$ (and electron
and quasihole operators). We conclude that all of the 
(topologically distinct) quasiparticle excitations are composites of the 
elementary quasiparticle corresponding to $\mathcal{O} = \psi 
e^{i\bar{\phi}/2\sqrt{2m+1/8}}$.

The charge of this excitation can be computed in many ways. One way, which 
doesn't involve too much formalism, is to consider $\mathcal{O}^2 = 
e^{i\bar{\phi}/\sqrt{2m+1/8}}$. An insertion of
$\mathcal{O}^2(w_0)$ into the correlation function (\ref{sumid}) 
gives the same correlation function back, but with an additional 
multiplicative factor of $\prod_i (\bar{w_i}-\bar{w_0})$. Thus this 
operator creates a Laughlin-like quasihole in the quasihole condensate.
Since the quasiholes are at filling $\nu_{anyon} = 1/(2m+1/8)$ and the 
quasiholes carry charge $e/4$, the charge accumulated at $w_0$ is 
\begin{displaymath}
\frac{e}{4} \cdot \frac{1}{2m+1/8} = \frac{2e}{16m+1}
\end{displaymath}
Dividing this result by $2$, we conclude 
that the charge associated with the elementary quasiparticle/quasihole 
is $\pm e/(16m+1)$.

As for the statistics, note that the two point correlator for $\mathcal{O}$ is
$\<\mathcal{O}(w_1)\mathcal{O}(w_2)\> = w_{12}^{-1}\bar{w}_{12}^{2/(16m+1)}$. 
The phase accumulated by this correlation function under a 
counter-clockwise exchange of $w_1, w_2$ is therefore 
$e^{i\theta} =  e^{i\pi(16m-1)/(16m+1)}$. Applying the conformal field 
theory conjectures, this is precisely the statistical (Berry) phase 
associated with a counter-clockwise quasiparticle exchange. 
Note, in particular, that the elementary quasiparticle is an 
\emph{abelian} anyon. This means that all the quasiparticles are abelian 
anyons (since they can be generated as composites of the elementary 
quasiparticle). Thus, the hierarchical state $\Psi^m_{hi}$ is 
actually an abelian quantum Hall state - even though the parent Pfaffian 
state is non-abelian!

This result is not as strange as it first appears. Indeed, as we mentioned 
earlier, previous analysis \cite{AK2006} has shown that an array of pinned  
quasiholes (or equivalently a quasihole Wigner crystal) can give rise to 
an abelian state (see also \cite{GS0603} and in the disordered case,\cite{RL0104}). 
It is therefore not surprising that a finite density of 
free quasiholes can give rise to abelian (hierarchical) states.

To complete our analysis of $\Psi^m_{hi}$, we compute its
thermal Hall conductance $K_H$. Our computation is based on the general
correspondence between the edge modes of a quantum Hall state and the 
modes of the conformal field theory used to define its wave 
function. \cite{W9505} Applying this to $\Psi^m_{hi}$, we see that the 
edge  contains one forward propagating majorana mode (corresponding to 
$\psi(z)$), one backward propagating majorana mode (corresponding to 
$\bar{\psi}(\bar{z})$), one forward propagating boson (corresponding to 
$\phi(z)$) and one backward propagating boson (corresponding to 
$\bar{\phi}(\bar{z})$). Recall that each chiral boson mode gives a 
contribution of $\pm \frac{\pi^2 k_B^2}{3h} T$ to the the thermal Hall 
conductance, while each majorana mode contributes half as much. The total 
thermal Hall conductance for $\Psi^m_{hi}$ is therefore $(1/2 - 1/2 + 1 - 
1) \frac{\pi^2 k_B^2}{3h} T = 0$.

\section{Relation to the Jain states and the strong pairing state}
\label{jain}
Since $\Psi^m_{hi}$ is abelian, it is natural to wonder how it relates to
the usual Jain state (or equivalently, hierarchical state \cite{BW9045}) 
at filling $\nu = 8m/(16m+1)$. To this end, note that the elementary 
quasiparticle/quasihole in the Jain state has charge $\pm e/(16m+1)$ and
statistical phase $e^{i\pi(16m-1)/(16m+1)}$. Similarly, note that 
the thermal Hall conductance of the Jain state is $8m$ (in units of 
$\frac{\pi^2 k_B^2}{3h} T$). Comparing with the results above, we see 
that while the quasiparticle charges and statistics of the two states are 
identical, their thermal Hall conductances are different. Thus 
$\Psi^m_{hi}$ is distinct from the $\nu = 8m/(16m+1)$ Jain state.

A simpler, but less fundamental distinction between the two states
can be obtained by examining their shifts $\mathcal{S}$ on the sphere. 
One can check that for the Jain state, the number of flux quanta $N_\phi$ 
is related to the number of electrons $N_e$ by 
\begin{equation}
N_{\phi} = \frac{16m+1}{8m}N_e - (8m+2)
\end{equation}
Thus the shift is $\mathcal{S} = 8m+2$. On the other 
hand, a simple calculation shows that the shift for $\Psi^m_{hi}$ is 
$\mathcal{S} = 5/2$. (One way to derive this is to note that the 
integral in (\ref{wf}) is only nonvanishing on the sphere if each quasihole 
coordinate $w_i$ occurs with the same maximum power as $\bar{w}_i$. This 
leads to the relation $N_e/2 = 2m(N_{qh}-1)$ where $N_e,N_{qh}$ are the 
number of electrons and quasiholes, respectively. Combining this with the 
relation between the number of flux quanta and number of quasiholes in the 
Pfaffian state, $N_{\phi} = 2 N_e - 3 + N_{qh}/2$, gives the required 
shift $\mathcal{S} = 5/2$). This difference in shifts provides a simple
way to distinguish the two states numerically.

While the $\Psi^m_{hi}$ are distinct from the Jain states and
the standard hierarchical states, we would like to mention that they appear
to be equivalent to hierarchical states derived from the strong-pairing $\nu = 
1/2$ state. \cite{H8375} Indeed, consider the strong-pairing 
$\nu = 1/2$ state, which can be thought of as a $\nu = 1/8$ Laughlin state 
of tightly bound pairs of electrons. The elementary quasihole in this state
carries charge $e/4$ and statistical phase $e^{i\pi/8}$. Following the 
usual abelian hierarchy construction,  \cite{H8305,H8483} 
one can consider Laughlin-like incompressible liquids formed out of these 
quasiholes. These states can 
occur at quasihole filling fractions $\nu_{qh} = 1/(2m+1/8)$ for any 
positive integer $m$. The corresponding electron filling fraction for 
these states is given by
\begin{equation}
\nu = \frac{1}{2} - \frac{1}{16} \cdot \nu_{qh} = \frac{8m}{16m+1}
\end{equation}
Thus, we see that these
states occur at the same filling fraction at $\Psi^m_{hi}$. In addition, 
one can check that the elementary quasiparticle/quasihole has charge 
$\pm e/(16m+1)$ and statistical phase $e^{i\pi(16m-1)/(16m+1)}$. Finally,
it is easy to see that the thermal Hall conductance of this state vanishes.
All of these properties agree exactly with the $\Psi^m_{hi}$. Therefore, it
appears that they carry the same topological order. We expect that they 
are in the same universality class - that is, one can go from one state to 
the other by continuously varying parameters in the Hamiltonian, without a 
phase transition.

\section{Quasiparticles vs. quasiholes}
\label{qpqh}
So far we have focused our attention on finding incompressible states 
formed out of quasiholes. However, it is equally natural to consider 
states formed out of \emph{quasiparticles}. In this section, we address 
this question. We construct a series of incompressible hierarchical states 
$\chi^m_{hi}$, analogous to $\Psi^m_{hi}$ but composed out of 
quasiparticles.

The first step is to construct incompressible states 
of point-like non-abelian anyons in a strong magnetic field. These anyons
should carry the statistics and charges of the quasiparticles in the 
Pfaffian state. 

Our approach, as before, is to construct lowest Landau level states with 
high order zeros when particles coincide. Since quasiparticles 
carry the opposite charge of quasiholes, these lowest Landau level wave 
functions are given by row vectors 
$(\tilde{\chi}_1, ... ,\tilde{\chi}_{2^{N-1}})$ of $2^{N-1}$ 
\emph{analytic} functions rather than anti-analytic functions. On the 
other hand, since quasiparticles have the same statistics as quasiholes, 
the $\tilde{\chi}_i$ must transform in the same way as before under particle 
exchange.

Previously, we constructed a series of lowest Landau level states with
the appropriate zero structure 
using the conformal blocks of a CFT correlator
\begin{displaymath}
\<\bar{\si}(\bar{w}_1)e^{i\sqrt{2m+\frac{1}{8}}\bar{\phi}(\bar{w}_1)} \cdots
\bar{\si}(\bar{w}_{2N})e^{i\sqrt{2m+\frac{1}{8}}\bar{\phi}(\bar{w}_{2N})}\>_\alpha
\end{displaymath}
Since in this case we need analytic functions rather than anti-analytic 
functions, it is tempting to use the same approach but with the correlator 
\begin{displaymath}
\<\si(w_1)e^{i\sqrt{2m+\frac{1}{8}}\phi(w_1)} \cdots
\si(w_{2N})e^{i\sqrt{2m+\frac{1}{8}}\phi({w}_{2N})}\>_\alpha
\end{displaymath}
Unfortunately, this 
construction does not work as the resulting collection of analytic 
functions $\tilde{\chi}^m_\alpha$ satisfy the opposite 
transformation law from what is required. 
That is, they transform as $\tilde{\chi}^m_{\alpha,ex} =\sum_\beta 
\tilde{\chi}^m_\beta U_{\beta \alpha}$ instead of 
$\tilde{\chi}^m_{\alpha,ex} 
=\sum_\beta \tilde{\chi}^m_\beta U_{\beta \alpha}^*$ (here $U_{\alpha 
\beta}$ is the unitary matrix associated with the particle exchange).

To construct a legitimate state, we need to modify the 
correlator so that it has the opposite monodromy under
particle exchange. To this end, we separate out the
abelian part of the correlator, and then regroup terms:

\begin{eqnarray}
\<\si(w_1)e^{i\sqrt{2m+\frac{1}{8}}\phi(w_1)} \cdots \>_\alpha &=& \prod_{i < j} 
w_{ij}^{2m+\frac{1}{8}} \<\si(w_1) \cdots \>_\alpha \nonumber \\
&=&  \left(\prod_{i < j} w_{ij}^{-\frac{1}{8}} \<\si(w_1) \cdots \>_\alpha \right) \nonumber \\
& \cdot &  \left(\prod_{i < j} w_{ij}^{2m+\frac{1}{4}} \right)  
\end{eqnarray}

We can reverse the monodromy of the second term by simply changing the
exponent $2m + 1/4 \rightarrow 2m - 1/4$. Reversing the monodromy of the
first term is more complicated. However, we can achieve this goal using 
the fact that the monodromy of this term is described by the spinor 
representation of $SO(2N)$. Indeed, this term transforms 
under a particle exchange $i \leftrightarrow j$, just like the $2^{N-1}$ basis 
vectors of the spinor representation transform under a $\pi/2$ rotation in 
the $ij$ plane. \cite{NW9629} Since the spinor representation is equivalent 
to its conjugate when $N$ is even, one can obtain an expression with the opposite 
monodromy by simply multiplying by the (unitary) matrix $R_{\alpha\beta}$ which describes 
this equivalence. In more detail: let
$R_{\alpha\beta}$ be the unique (up to phase) unitary matrix with the
property that $R^{-1}_{\alpha \beta} U_{\beta \gamma} R_{\gamma \delta} = U^*_{\alpha 
\delta}$ for all the unitary matrices $U$ in the spinor representation of $SO(2N)$.
Then, $\prod_{i < j} w_{ij}^{-1/8} \<\si(w_1) \cdots\>_\beta R_{\beta \alpha}$
has the opposite monodromy from $\prod_{i < j} w_{ij}^{-1/8} \<\si(w_1) \cdots\>_\alpha$
under particle exchange. One can give an explicit form for $R$ if one labels the 
conformal blocks by $\alpha = (\alpha_1,...\alpha_{N-1})$ where 
$\alpha_i = 1,2$ depending on whether particles $2i-1,2i$ are in the 
$1,\psi$ fusion channel. In that basis, $R$ can be written as a product of 
Pauli matrices, $R = \si^y_1 \si^x_2 \si^y_3 \si^x_4 ...\si^y_{N-1}$. \cite{Georgi99}

Making these two modifications, and recombining the two terms, we 
arrive at the following candidate states:

\begin{equation}
\tilde{\chi}_{\alpha}^m(w_1,...,w_{2N}) =
\<\si(w_1)e^{i\sqrt{2m-\frac{3}{8}}\phi(w_1)}\cdots\>_{\beta} R_{\beta \alpha}
\end{equation}

By construction, the $\tilde{\chi}_\alpha^m$ are analytic and transform in 
the appropriate way under particle exchange. Thus, they describe 
legitimate lowest Landau level states.



As an example, consider the case $N = 2$. In that case, the
$\tilde{\chi}_\alpha^m$ are given (up to constant factor) by
\begin{eqnarray}
\tilde{\chi}_{1,2}^m(w_1,w_2,w_3,w_4) &=&
\pm (\sqrt{w_{13}w_{24}} \mp
\sqrt{w_{14}w_{23}})^{\frac{1}{2}} \nonumber \\
& \cdot & \prod_{i<j} w_{ij}^{2m-\frac{1}{2}}
\end{eqnarray}
where $w_{ij} = w_i - w_j$. One can see that when particles
$1$ and $2$ are exchanged, $(\tilde{\chi}_1 , \tilde{\chi}_2) \rightarrow
(\tilde{\chi}_1 ,- i \tilde{\chi}_2)$. This is exactly the required unitary
transformation.

In addition to being in the lowest Landau level, the $\chi^m$ satisfy 
our condition of having high order zeros when two particles approach each 
other. Thus, they appear to be good candidate ground states for a model with 
short-range, two-body, repulsive interactions. As before, one can complete 
the picture by constructing model interactions for which they are the exact 
ground state.

Thus, just like the $\Psi^m$, the $\chi^m$ suggest a series of
hierarchical states $\chi^m_{hi}$ that can arise from the Pfaffian state. 
These states will occur if there is a finite density of quasiparticles, and
the quasiparticles form one of the $\chi^m$ states. 

What are the properties of these states? Let us begin with the filling
fraction. Since the anyon filling fraction for $\chi^m$ is 
$\nu_{anyon} = 1/(2m-3/8)$, the electron filling fraction for $\chi^m_{hi}$ is 
\begin{equation}
\nu = \frac{1}{2} + \frac{1}{16} \cdot \nu_{anyon} = \frac{8m-1}{16m-3}
\end{equation}
In particular, the simplest hierarchical state (corresponding to $m=1$) 
occurs at $\nu = 7/13$.

Next, we write down the wave functions for these states in terms of the
electron coordinates, $\{z\} = \{z_1,...z_{2M}\}$. We have
\begin{equation}
\chi^m_{hi}(\{z\}) = \int dw \sum_\alpha
\chi^m_\alpha(\{w\})
\Phi_{\alpha,qp}(\{z\};\{w\})
\label{wf2}
\end{equation}
where $\chi^m_{\alpha} = \tilde{\chi}^m_{\alpha}
\cdot e^{1/16 l^2 \sum_i |w_i|^2}$ are the quasiparticle
wave functions, and $\Phi_{\alpha,qp} = \tilde{\Phi}_{\alpha,qp} \cdot 
e^{-1/4 l^2 \sum_i |z_i|^2}$ are the electron wave functions for fixed
quasiparticle positions $\{w\}$.

Unlike the quasihole case, there is no canonical form for these
quasiparticle wave functions. We will use the wave function
\begin{align}
&\tilde{\Phi}_{\alpha,qp}(\{z\};\{w\}) = \nonumber \\
&\<\si'(w_1)e^{-\frac{i}{\sqrt{8}}\phi'(w_1)} \cdots
\psi'(z_1)e^{i\sqrt{2}\phi'(z_1)} \cdots\>_{\alpha}
\end{align}
as it is particularly convenient for our analysis. (Note that this
wave function has unphysical singularities as $w_i \rightarrow z_j$.
Strictly speaking these singularities need to be regularized in some
way. However, we will ignore this regularization as it plays no
role in the universal long distance structure of the wave function).

We next make use of the identity
\begin{align}
&\<\si(w_1) \cdots \>_\alpha \<\si'(w_1) \cdots \psi'(z_1) \cdots \>_\beta 
R_{\alpha \beta} \nonumber \\
&= \<e^{\frac{i}{2}\phi''(w_1)} \cdots \cos(\phi''(z_1)) \cdots \>
\label{sumid2}
\end{align}
(up to a constant factor) where $\phi''$ is a free boson field. 
(A formal justification of this identity can be found in \cite{DVV8985} in the 
second paragraph after equation (7.32). In addition, we have explicitly 
verified the relation for $N=0,1,2$). Applying this identity, the 
sum in (\ref{wf2}) can be simplified to
\begin{align}
&\sum_\alpha \chi^m_\alpha(\{w\}) \Phi_{\alpha,qp}(\{z\};\{w\}) = \nonumber \\
& \<e^{i\sqrt{2m-\frac{3}{8}}\phi(w_1)-\frac{i}{\sqrt{8}}\phi'(w_1) + \frac{i}{2}\phi''(w_1)} \cdots \nonumber \\
& e^{i\sqrt{2}\phi'(z_1)}\cos(\phi''(z_1)) \cdots\> 
\cdot e^{-\frac{1}{4 l^2} \sum_i |z_i|^2 -\frac{1}{16 l^2} \sum_i |w_i|^2} 
\end{align}

We have now expressed $\chi^m_{hi}$ in terms of a single CFT correlator and  
are therefore in a position to use the CFT approach to compute its 
properties. A calculation analogous to the one for $\Psi^m_{hi}$ shows that 
all the quasiparticle excitations are composites of the elementary 
quasiparticle corresponding to $\mathcal{O} = 
e^{i\phi/2\sqrt{2m-3/8}-i\phi''}$. It is also straightforward to show that 
the elementary quasiparticle/quasihole carries charge $\pm e/(16m-3)$ and 
exchange statistics $e^{i\theta} = e^{i\pi(16m-1)/(16m-3)}$.
Thus, $\chi^m_{hi}$ is an \emph{abelian} state - just like $\Psi^m_{hi}$. 
As for the thermal Hall conductance, note that the edge contains three 
forward propagating chiral bosons (corresponding to $\phi,\phi',\phi''$ so 
that the total thermal Hall conductance is $1+1+1 = 3$ (in units of 
$\frac{\pi^2 k_B^2}{3h} T$).

How does this state relate to the Jain state at filling $\nu = 
(8m-1)/(16m-3)$? A simple calculation shows that the elementary 
quasiparticle in the Jain state has the same charge and statistics
as $\chi^m_{hi}$. Nevertheless the two states are distinct as they have 
different thermal Hall conductances: the thermal Hall conductance for the 
Jain state is $3 - 8m$ (in appropriate units) - rather than $3$. In 
addition, while the shift for the Jain state is $\mathcal{S} = 3-8m$, the
shift for $\chi^m_{hi}$ is $\mathcal{S} = (28m-3)/(8m-1)$. This difference 
in shifts provides a simple way to distinguish the two states numerically.

While the $\chi^m_{hi}$ are distinct from the Jain states, they appear to
be equivalent to hierarchical states derived from the $331$ $\nu = 1/2$ 
state. \cite{H8375,HR8886} Indeed, the elementary quasiparticle in the $331$ state carries 
charge $e/4$ and statistical phase $e^{3i\pi/8}$. Following the usual 
hierarchical construction,  \cite{H8305,H8483} these quasiholes can form Laughlin-like 
incompressible liquids at quasiparticle filling fractions $\nu_{qp} = 
1/(2m-3/8)$. The corresponding electron filling fraction for these states
is 
\begin{equation}
\nu = \frac{1}{2} + \frac{1}{16} \cdot \nu_{qp} = \frac{8m-1}{16m-3}
\end{equation}
Thus, these
states occur at the same filling fraction as $\chi^m_{hi}$. In addition,
one can check that the elementary quasiparticle/quasihole has charge
$\pm e/(16m-3)$ and statistical phase $e^{i\pi(16m-1)/(16m-3)}$. Finally,
it is easy to see that the thermal Hall conductance of this state is $3$
in appropriate units. All of these properties agree exactly with the 
$\chi^m_{hi}$. Therefore, it appears that they carry the same topological 
order. We expect that they are in the same universality class - that is, 
one can go from one state to the other by continuously varying parameters 
in the Hamiltonian, without a phase transition.

\section{Higher levels in the hierarchy}
\label{hihi}
As in the original abelian hierarchy, the states $\Psi^m_{hi}$, 
$\chi^m_{hi}$ may give rise to daughter states with filling fractions near 
$8m/(16m+1)$ and $(8m-1)/(16m-3)$ respectively. These second level 
hierarchical states may then give rise to their own daughter states and so 
on. In this way, the Pfaffian state may give rise to an infinite hierarchy 
of quantum Hall states. We will not describe this hierarchy in detail, 
since the parent states $\Psi^m_{hi}$, $\chi^m_{hi}$ are abelian and 
therefore the analysis is very similar to the original abelian hierarchy. \cite{H8305,H8483}
Nevertheless, with very little work, we can say quite a bit about these 
higher level hierarchical states. 

The key point is that $\Psi^m_{hi}$ and $\chi^m_{hi}$ have the same 
quasiparticle charges and statistics as the usual abelian hierarchical 
states at the corresponding filling fraction. Since these quantities are 
the only ones that enter in the hierarchical construction, it follows that 
the descendant states of $\Psi^m_{hi}$, $\chi^m_{hi}$ must also have the 
same filling fractions, quasiparticle charges, and quasiparticle 
statistics as the usual abelian hierarchical states. (Despite this 
similarity, these states are distinct from the usual abelian hierarchical 
states: just as $\Psi^m_{hi}$, $\chi^m_{hi}$ have a different thermal Hall 
conductance from the corresponding abelian hierarchical states, this will 
also hold for all descendant states). 

This result allows us to compute the universal properties of the higher 
level states very easily. In particular, we can find the filling fractions 
for these states. One finds that the hierarchy gives rise to all odd 
denominator fractions in the range $15/32 < \nu < 13/24$ and no others. 
(Here, we have assumed, as in the original hierarchy construction, that the 
descendant states of $\Psi^m_{hi}$, $\chi^m_{hi}$ are constructed out of 
Laughlin-like states of \emph{elementary} quasiparticles or quasiholes - 
that is quasiparticles or quasiholes with the minimal charge. If instead, 
the quasiparticles cluster into higher charges which then form 
Laughlin-like states, other filling fractions can be realized). We would 
like to mention though that unlike the usual hierarchy, one can obtain 
obtain multiple states at the same filling fraction. For example, two $\nu 
= 16/33$ states can be obtained: one as the $\Psi^2_{hi}$ state and one as 
a descendant of the $8/17$ state. These two $\nu = 16/33$ states have the 
same quasiparticle statistics and charges, but different thermal Hall 
conductances.

\section{Anti-Pfaffian vs. Pfaffian}
\label{antipf}
The discussion so far has focused on hierarchical states derived from the 
Pfaffian states. However, since the particle-hole conjugate of the Pfaffian 
(or ``anti-Pfaffian" \cite{LHR0706, LRNF0707}) is an equally strong 
candidate for the observed $\nu = 5/2$ plateau, it is worth applying the 
analysis to this state as well. 

The simplest way to proceed is to note that the states derived from the 
anti-Pfaffian are just the particle-hole conjugates of those derived from 
the Pfaffian. It follows that these states - which we will call 
$\Psi^m_{apf,hi}$, $\chi^m_{apf,hi}$ - are necessarily abelian and have 
filling fractions $(8m+1)/(16m+1)$ and $(8m-2)/(16m-3)$ respectively. The 
simplest states (with $m=1$) occur at $9/17$ and $6/13$. 

If one considers higher levels in the hierarchy as above, one can 
construct a state with any odd denominator filling fraction $\nu$ with 
$11/24 < \nu < 17/32$. As in the case of the Pfaffian, all of these states 
have the same quasiparticle charges and statistics as usual abelian 
hierarchical states at the same filling fraction. However, they are 
distinct from the usual abelian hierarchical states - as they have 
different thermal Hall conductance.

While these states are distinct from the usual abelian hierarchical 
states, it is not clear that they are distinct from the hierarchical 
states derived from the Pfaffian state. For example, the $\nu = 8/17$ 
state derived from the anti-Pfaffian $6/13$ state appears to have all the 
same properties as the Pfaffian derived $\nu = 8/17$ state. Specifically, 
the two states have the same quasiparticle charges, statistics, \emph{and} 
the same thermal Hall conductance. 

\section{Candidate states for a quasihole Wigner crystal}
\label{locqh}
In this section we consider the problem of a quasihole Wigner crystal. We 
construct two candidate states for this system using an approach similar 
to the incompressible liquid case.

More specifically, the problem we wish to consider is the following. 
Take a FQH system whose ground state at half filling is the Pfaffian 
state, and suppose that the filling is slightly less than $1/2$ so that 
there is a finite density of quasiholes. Suppose further that these 
quasiholes form a Wigner crystal (e.g. a triangular lattice). For 
conceptual simplicity, assume the quasiholes are pinned at the
lattice sites so that there are no phonon modes at low energies.
Because of the quasiholes' non-abelian statistics, there will be $2^{N-1}$ 
nearly degenerate states for a lattice of $2N$ quasiholes. While the
splitting between these states vanishes in the limit of infinitely large 
lattice spacing, it will be nonzero for any finite sized lattice. Thus, a 
particular linear combination of the $2^{N-1}$ low lying states
will be selected as the ground state. We would like to understand what 
kind of ground states can occur in this system and what 
properties the associated FQH states have - e.g. quasiparticle 
statistics, thermal Hall conductance, etc. 

As before, our approach will be to propose candidate ground states for 
this system and then to analyze their properties. Such ground states are 
specified by vectors with $2^{N-1}$ components. As before, the CFT ansatz 
suggests two natural vectors with $2^{N-1}$ components. The first is 
defined by
\begin{equation}
\Psi_\alpha =
\<\bar{\si}(\bar{w}_1) \cdots\ \bar{\si}(\bar{w}_{2N})\>_{\alpha}
\end{equation}
where $w_1,...w_{2N}$ are the positions of the pinning sites (which we 
have assumed form a triangular lattice). This state is closely related to 
the series of hierarchical states $\Psi^m$. Similarly, the second 
candidate state is closely related to the series of hierarchical states 
$\chi^m$. This state is given by
\begin{equation}
\chi_{\alpha} =
\<\si(w_1)\cdots \si(w_{2N})\>_{\beta} R_{\beta \alpha}
\end{equation}

Unlike the hierarchical states $\Psi^m$, $\chi^m$ we do not have any 
arguments for why these states may be favored energetically. In fact we do 
not know any Hamiltonian for which they are the exact ground state. 
Nevertheless, we will assume that they are physical ground states and see 
what their properties are.

To derive these properties, we write out the wave functions for $\Psi, 
\chi$ in terms of the electron coordinates $\{z\}$. We have 
\begin{eqnarray}
\Psi(\{z\}) &=& \sum_\alpha \Psi_\alpha \cdot \Phi_{\alpha,qh}(\{z\}) \nonumber \\
& = & \<\si(w_1,\bar{w}_1)e^{\frac{i}{\sqrt{8}}\phi(w_1)}\cdots 
\psi(z_1)e^{i\sqrt{2}\phi(z_1)} 
\cdots\> \nonumber \\
&\cdot& e^{-\frac{1}{4 l^2} \sum_i |z_i|^2}
\end{eqnarray}
where the second equality follows from the same reasoning as (\ref{sumid}).
Similarly, one can show that $\chi(\{z\})$ is given by the correlator
\begin{align}
& \chi(\{z\}) = \<e^{\frac{i}{\sqrt{8}}\phi'(w_1) + \frac{i}{2}\phi''(w_1)}
\cdots \nonumber \\
& e^{i\sqrt{2}\phi'(z_1)}\cos(\phi''(z_1)) \cdots \> \cdot e^{-\frac{1}{4 l^2} \sum_i |z_i|^2}
\end{align}

To proceed further, we make another bold assumption: we assume that the 
CFT approach is still applicable even though the quasiholes are 
fixed in space and are not free to move as in a normal hierarchical state.
That is, we assume that quasiparticle excitations can still be constructed
by inserting operators $\mathcal{O}$ into the above correlators, and that 
the statistics of these quasiparticles can still be computed from the 
monodromy of these correlators. We would like to emphasize that this is a 
conjecture - and is on even less firm ground than usual applications of
the CFT approach. Indeed, the validity of the CFT approach in this 
context likely depends on the specific choice of lattice. For example, one 
can imagine that on some lattices (particularly those with an even number 
of quasiholes per unit cell) the wave functions $\Psi, \chi$ may describe 
highly dimerized states where quasiholes pair up with a fixed partner in 
either the $1$ or $\psi$ fusion channel. In this case, the CFT approach 
could break down in the same way that it fails for the Wigner crystal 
states that occur at large $m$ in the usual Laughlin series. By using
the CFT approach we are implicitly assuming that this kind of 
dimerization scenario does not occur - at least on the triangular lattice.

We begin with the state $\Psi$. The allowed operators $\mathcal{O}$ are 
those which are local with respect to the electron operator $\psi 
e^{i\sqrt{2}\phi}$. A complete list of such operators (or more accurately 
equivalence classes of such operators) is given by: $\{\upsilon 
e^{in\phi/\sqrt{2}}\}$ where $\upsilon = 1, \si(z,\bar{z}), \mu(z,\bar{z}), 
\psi$ and $n$ is integer or half integer depending on whether $\upsilon = 
1, \psi$ or $\upsilon = \si(z,\bar{z}), \mu(z,\bar{z})$. One can check 
that all of these can be generated from the elementary operator $\mathcal{O} = 
\si(z,\bar{z})e^{i\phi/2\sqrt{2}}$ (and the electron operator). Thus, all 
the quasiparticles are composites of an elementary quasiparticle 
corresponding to $\mathcal{O}$.

Using the usual arguments, it is easy to check that the elementary 
quasiparticle carries charge $\pm e/4$, and statistical phase $e^{i\pi/8}$.
In particular, the state $\Psi$ is \emph{abelian}.
One can go further and compute the thermal Hall conductance. Since the CFT 
has one forward propagating majorana mode, one backward propagating 
majorana mode and one forward propagating boson, the thermal Hall 
conductance is $1/2 - 1/2 +1 = 1$ in appropriate units. On the other hand,
the electric Hall conductance is $1/2$ (like the parent Pfaffian state)
since the quasiholes are all localized.

We would like to mention that these properties agree exactly with the 
strong pairing $\nu = 1/2$ state. \cite{H8375} Thus, it appears that $\Psi$ is in the
same universality class as the strong pairing state.

Next, consider the state $\chi$. In this case, one finds that all allowed 
operators are composites of the elementary operator $\mathcal{O} = 
e^{i(\sqrt{1/8}\phi'(w_1) + \phi''(w_1)/2)}$. One finds that the
corresponding elementary quasiparticle/quasihole carries charge $\pm e/4$ 
and statistical phase $e^{3i\pi/8}$. The thermal Hall conductance is $1+1 = 
2$ in appropriate units. Comparing with the $331$ $\nu = 1/2$ state \cite{H8375, HR8886} 
we conclude that $\chi$ is in the same universality class as this state.
 
Putting this all together we conclude that a lattice of localized 
quasiholes can naturally give rise to abelian states. We have found two 
such candidates, $\Psi$, and $\chi$ - with different quasiparticle 
statistics and thermal Hall conductance. 

As we mentioned earlier, these candidate states agree well with Kitaev's 
study of quasihole lattices \cite{AK2006} (see also \cite{GS0603} and in the disordered 
case, \cite{RL0104}). In that work, the author described an exact solution of a 
triangular lattice of quasiholes with a nearest neighbor interaction that favored the $1$ 
fusion channel. His conclusion was that the ground state was abelian and in the strong 
pairing phase - the same universality class as $\Psi$. One can also 
consider the same model but with a nearest neighbor interaction that 
favors the $\psi$ fusion channel. Using the same approach as in 
\cite{AK2006}, one finds that the ground state is again abelian. Moreover, 
one can show that it is in the $331$ phase - the same universality class as 
$\chi$. This comparison suggests that our candidate states - and our 
construction in general - is at least somewhat natural.

\section{Conclusion}

In this paper, we have investigated collective states that can arise
from a finite density of non-abelian $e/4$ quasiholes in a magnetic field. 
We have focused on two types of collective states: 
incompressible liquids and Wigner crystals. In the incompressible liquid 
case, we have proposed a natural series of incompressible states 
$\Psi^m$. These states are good candidate ground states for a model 
where the quasiholes have short-range, repulsive, two-body interactions
and are in their lowest effective Landau level.
The $\Psi^m$ are associated with hierarchical FQH states
$\Psi^m_{hi}$ derived from the Pfaffian state. Interestingly, these 
hierarchical states - which occur at filling fraction $\nu = 
8m/(16m+1)$ - are \emph{abelian} quantum Hall states. We have also 
investigated incompressible liquids of $e/4$ quasiparticles. In that case, 
we have proposed another series of incompressible states $\chi^m$. The 
resulting hierarchical states $\chi^m_{hi}$ are again abelian and occur 
at filling fraction $\nu = (8m-1)/(16m-3)$.

In the Wigner crystal case, we have proposed two candidate ground 
states $\Psi$,$\chi$, closely related to the incompressible liquids 
$\Psi^m$, $\chi^m$. We have analyze the properties of these states and 
we have shown that these states are also abelian. Our results can be 
compared with those obtained from a microscopic model \cite{AK2006} of a 
quasihole lattice. It appears that the microscopic analysis agrees with the
results presented here - and suggests the same two abelian phases.

While all of the states we have constructed are abelian,
we would like to reiterate that this is not the only possibility. A 
finite density of non-abelian quasiparticles need not always destroy the
non-abelian statistics in the Pfaffian state. As we discussed earlier, 
one can imagine a scenario as in \cite{BS0823} where the 
quasiparticles pair and form tightly bound charge $e/2$ abelian 
anyon molecules which in turn form a Laughlin-like state. The result 
is a hierarchical state with the same non-abelian statistics as the Pfaffian 
state. One can also consider a similar scenario in the Wigner crystal case. 
Depending on the interactions, the quasihole lattice may dimerize, with 
pairs of quasiholes favoring the $1$ (or $\psi$) fusion channel. Again, the 
result is a non-abelian state (in fact, in the same universality class as the 
original Pfaffian state). One can consider this to be a crude rule of thumb: 
hierarchical or Wigner crystal states composed out of $e/4$ quasiparticles
are non-abelian if the quasiparticles pair, and abelian otherwise.

A natural direction for future research would be to extend 
our analysis to general Read-Rezayi states. It would be particularly 
interesting to apply these methods to the problem of a Wigner crystal of 
Read-Rezayi quasiholes. Indeed, unlike the Pfaffian case, 
microscopic models of this system have not been solved exactly
(except in the case of a one dimensional chain of quasiholes 
\cite{FTLT0709}). The approach outlined in this paper could suggest 
potential phases of this poorly understood system.

\acknowledgments
We would like to thank Ady Stern and Chetan Nayak for useful discussions.
This work was supported by the Microsoft Corporation, by NSF grants DMR-05-41988 and DMR-05-29399, and by the Harvard Society of Fellows.

\newcommand{\noopsort}[1]{} \newcommand{\printfirst}[2]{#1}
  \newcommand{\singleletter}[1]{#1} \newcommand{\switchargs}[2]{#2#1}

\end{document}